\documentclass[epj,nopacs]{svjour}
\usepackage{graphicx}
\begin{document}
\title{Phenomenological model for the $\overline{K}N\to K\Xi$ reaction}
\author{D.A.\ Sharov\thanks{sharov@depni.sinp.msu.ru} \and V.L.\ Korotkikh\thanks{vlk@lav1.sinp.msu.ru} \and D.E.\ Lanskoy\thanks{lanskoy@sinp.msu.ru}}
%
%
\institute{Institute of Nuclear
Physics, Moscow State University, 119991 Moscow, Russia}
\date{}
%
\abstract{
A phenomenological model for the $\overline{K}N\to K\Xi$ reaction is suggested. The model includes $s$ and $u$ channel exchanges by $\Lambda$, $\Sigma$, $\Sigma(1385)$, and $\Lambda(1520)$ and $s$ channel exchanges by above-threshold hyperonic resonances. Explicit expression for the propagator for a particle with spin $7/2$ is presented. High-mass and high-spin resonances play a significant role in the process. We deal with the whole set of existing experimental data on the cross sections and polarizations in the energy range from the threshold to 2.8 GeV in the center-of-mass system and reach a good agreement with experiments. Applications of the model to other elementary reactions of $\Xi$ production and to $\Xi$ hypernuclear spectroscopy are briefly discussed.}

\maketitle
\section{Introduction}
\label{s1}
Whereas $\Xi$ hyperons were discovered soon after $\Lambda$ and $\Sigma$ hyperons, our knowledge in the $S=-2$ sector remains much more limited. $\Xi$ production reactions are typically more complicated than those for $\Lambda$ or $\Sigma$ production, and their cross sections are small. Production of $\Xi$ in a binary process is possible with kaonic beams ($\overline{K}N\to K\Xi$), which are relatively less intense, while other projectiles, such as pions, photons, electrons, protons, inevitably lead to three- or even four-body final states. That is why the $\Xi$ production dynamics is still poorly studied.

Modern experimental facilities providing intense beams open new possibilities for systematic studies of $\Xi$ production. Photoproduction $\gamma p\to K^+K^+\Xi^-$ has been investigated at JLab by CLAS collaboration \cite{Price,Guo}. Higher statistics will become available after completing of data analysis of the new run g12 \cite{g12,Goe}. The intense kaonic beam is expected at J-PARC where $\Xi$ production from nuclear targets is announced as the highest priority task \cite{Nagae,Nagae2}. The elementary reaction $\overline{K}N\to K\Xi$ with formation of ground as well as resonant $\Xi$ states can be also studied at J-PARC \cite{Ahn}. PANDA collaboration proposed a study of the $\overline{p}p\to\overline{\Xi}\Xi$ reaction in view of the FAIR project \cite{PANDA,PANDA2}. Therefore, one may anticipate that much more empirical information on $\Xi$ ground state production as well as on the spectrum of $\Xi$ resonances will become available soon. Note that so far the spin and parity are known for the ground (octet), first excited (decuplet), and only one more resonant $\Xi$ state. Total number of observed $\Xi$ states is 11 (and not all of them are established with confidence) \cite{PDG}, which is much less than for $S=0$ and $S=-1$, and than predicted by quark models.

The $\overline{K}N\to K\Xi$ reaction was studied experimentally in the sixties and seventies \cite{Bell,Berg,Brie,Brie2,Burg,Carl,Carm,Daub,Gris,Haque,London,Rader,Trip,Tro,Sche} when the bubble chamber technique was popular. Later, the reaction left the focus of experimentalists, and only one measurement of the forward cross section at one energy was performed at KEK \cite{Iij}. Few related theoretical papers ({\it e.g.}, \cite{theor,theor2,theor3,theor4}) were published also in the past decades and dealt with small portions of the existing data. In these papers, some interesting features were pointed out, but no systematic theoretical picture has been ever presented.

Later, closely related problems of $\Xi N$ interaction and $\Xi$ hypernuclear dynamics attracted some attention. Data on the $\overline{K}N\to K\Xi$ reaction were discussed \cite{DG} in view of formation of $\Xi$ hypernuclei by the $(\overline{K},K)$ reaction from nuclei. The $(K^-,K^+)$ reaction on nuclei was studied theoretically ({\it e.g.}, \cite{DG,Mot,XiHyp,XiHyp2}). The forward cross section of the elementary $K^-p\to K^+\Xi^-$ process needed in the calculations was taken from empirical data.

The $\Xi N$ interaction was analyzed theoretically within unitary meson-exchange models of baryon-baryon interactions extended to the full baryonic octet \cite{OBE1,OBE2,OBE3}. The $\Xi N$ potentials derived from those models can be discriminated, in principle, by the comparison with $\Xi$ hypernuclear data. However, such data \cite{DG,Fuk,Kh,Naka} are rather scarce and ambiguous now, therefore, the theoretical parameters (particularly, $\Xi$ coupling constants) relevant to the $S=-2$ sector, are not yet approved empirically. Here, the first attempt to extract the $\Xi N$ interaction cross section from analysis of the rescattering of $\Xi$ produced inside a nucleus \cite{Tama} should be mentioned.

Data on photoproduction reaction $\gamma p\to K^+K^+\Xi^-$ \cite{Price} were analyzed theoretically by Nakayama {\it et al.} \cite{Nak} in a diagram approach with coupling constants for $\Xi$ derived from flavor SU(3) symmetry. Probably, this is the first direct check of consistence of deduced $\Xi$ coupling constants with data.

In this paper, we present a systematical analysis of the existing data on the $\overline{K}N\to K\Xi$ reaction at c.m. energy $\sqrt{s}<2.8$ GeV in various charge channels. We develop a phenomenological model, which describes all the available data fairly well. Our analysis discloses some features of the $\Xi$ production dynamics and also shows restrictions caused by the limited amount of the existing data. Some preliminary results have been published in conference proceedings \cite{KLSh}. The information extracted from the simplest binary reaction $\overline{K}N\to K\Xi$ can be helpful for further studies of more complicated reactions like $\gamma p\to K^+K^+\Xi^-$.

In sect.\ \ref{s2}, we present our model. The experimental data are introduced  in sect.\ \ref{s3}. Our results are presented and discussed in sect.\ \ref{s4}. We finish with conclusion and some outlook.

\section{Model}
\label{s2}
\begin{figure}
\includegraphics*{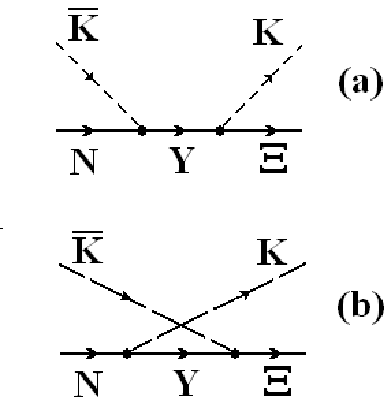}

\caption{Diagrams for the $s$ (a) and $u$ (b) channel exchanges.}
\label{f1}       
\end{figure}

We adopt the standard diagram approach. The simplest diagrams for our process are presented in fig.\ \ref{f1}. We stress that the one-meson $t$ channel exchange is absent since no (nonexotic) double-strangeness meson exists. So only two-meson exchange is possible in the $t$ channel. Thus, we deal with the $s$ channel [fig.\ \ref{f1}(a)] and $u$ channel [fig.\ \ref{f1}(b)] exchanges giving some comments on the $t$ channel exchange in subsect.\ \ref{s4a}.

Our treatment is restricted to the Born approximation, which means that neither channel coupling nor initial and final state interaction is taken into account. More comprehensive approaches incorporating the above-mentioned effects and the unitary condition are known in the literature. They are reviewed, for instance, in \cite{Oller} for kaonic reactions and in \cite{Matsu} for pion photoproduction. We use here the simpler approach considering that this is adequate for the first systematic study of reaction $\overline{K}N\to K\Xi$.

We consider three charge channels of the reaction
\begin{eqnarray}
K^-p&\to &K^+\Xi^-,\label{e1}\\
K^-p&\to &K^0\Xi^0,\label{e2}
\end{eqnarray}
and
\begin{equation}
K^-n\to K^0\Xi^-,\label{e3}
\end{equation}
for which experimental data are available. Due to isospin relations, the three corresponding amplitudes are not independent:
\begin{eqnarray}
&M(K^-n\to K^0\Xi^-)&\nonumber\\ =&M(K^-p\to K^+\Xi^-)&+M(K^-p\to K^0\Xi^0).\label{e4}
\end{eqnarray}

Note that $\Sigma$ exchange is allowed for all the reactions in the $s$ as well as $u$ channel while $\Lambda$ exchanges is impossible in the $s$ ($u$) channel for the third (second) reaction.

Effective Lagrangian for $B(1/2^+)Y(1/2)K(0^-)$ vertices is chosen with pseudovector couplings (Hereafter, notation and conventions of \cite{BD} are adopted):
\begin{equation}
L=\frac{f_{BYK}}{m_\pi}\overline{B}\gamma_\mu\left\{\gamma_5\atop 1\right\} Y\partial^\mu K + h.c.,
\end{equation}
where the upper (lower) symbol in the curly brackets stands for positive (negative) parity of $Y$, respectively, other symbols have their usual meanings. Propagators for $1/2$ baryons have the standard form:
\begin{equation}
S(q)=\frac{\not q+M}{q^2-M^2}.
\end{equation}

For $B(1/2^+)Y(3/2)K(0^-)$ couplings, the form adopted is also conventional:
\begin{equation}
L=\frac{f_{BYK}}{m_\pi}\overline{B}\left\{1\atop\gamma_5\right\} Y_\mu\partial^\mu K + h.c.
\end{equation}

As for the $3/2$ propagator, several different prescriptions are known \cite{Wil,Ad,Ben}. Their fundamental advantages and drawbacks were discussed numerously ({\it e.g.}, \cite{P32,P322,P323}). Treating our model as a phenomenological one, we do not enter into this discussion. For $u$ channel exchanges as well as for $s$ channel exchanges by particles below the reaction threshold, we employ the propagator, which is shown \cite{Ben} to be the proper inverse of the Rarita-Schwinger equation:
\begin{eqnarray}
S_{\mu\nu}(q)=\frac{\not q+M}{3(q^2-M^2+iM\Gamma)}\nonumber\\ \times\left(3g_{\mu\nu}-\gamma_\mu\gamma_\nu -\frac{2q_\mu q_\nu}{M^2}-\frac{\gamma_\mu q_\nu-\gamma_\nu q_\mu}{M}\right).
\end{eqnarray}
Possible dependence of resonance widths on energy is neglected.

For $s$ channel diagrams with above-threshold intermediate particles, we choose the propagators according to the simple criterion as follows. Let us consider an isolated $s$ channel diagram [fig.\ \ref{f1}(a)], where $Y$ is a spin-3/2 resonance lying above the $K\Xi$ threshold and decaying into $K\Xi$. We require that the c.m. angular distribution of the products corresponds to the proper Legendre polynomial [$P_1(\cos(\theta))$ for $3/2^+$ and $P_2(\cos(\theta))$ for $3/2^-$ for the spin-independent part of the amplitude and the corresponding associated Legendre polynomials for the spin-flip part] not only at the nominal resonance mass, but also at the slopes of the resonant peak. We checked that the only propagator obeying this condition is one proposed by Adelseck {\it et al.} \cite{Ad}:
\begin{eqnarray}
S_{\mu\nu}(q)=\frac{\not q+\sqrt{s}}{3(q^2-M^2+iM\Gamma)}\nonumber\\ \times\left(3g_{\mu\nu}-\gamma_\mu\gamma_\nu -\frac{2q_\mu q_\nu}{s}-\frac{\gamma_\mu q_\nu-\gamma_\nu q_\mu}{\sqrt{s}}\right).
\end{eqnarray}
The same criterion is applied for higher spin resonances.

A lot of $Y$ resonances is known in the relevant energy range. PDG compilation \cite{PDG} gives eight resonances of four- and three-star status with $1.89<M<2.35$ GeV. On the other hand, branching ratios of $K\Xi$ decay are determined for none of them. Only for two of them, upper limits [3\% for $\Lambda(2100)$ and 2\% for $\Sigma(2030)$] are deduced \cite{PDG}.

Probably, the branching ratios are small for all the resonances, while the main decay channels are $\pi\Lambda$ (for $\Sigma$ resonances), $\pi\Sigma$, and $\overline{K}N$. This is physically natural since, first, the $K\Xi$ decay requires creation of an additional $\overline{s}s$ pair and, second, the resonances lie not far from the $K\Xi$ channel threshold. On the other hand, cross sections of the $\overline{K}N\to K\Xi$ reaction are also not large, so even small branching ratios can contribute sizably to the reaction. Thus, the role of the above-threshold resonances should be studied. Most of these resonances has high spins. Therefore, we need technicalities for 5/2 and 7/2 spins of exchanged particles.

For a $5/2$ spin, we use the effective Lagrangian
\begin{equation}
L=\frac{f_{BYK}}{m_\pi^2}\overline{B}\left\{\gamma_5\atop 1\right\} Y_{\mu\nu}\partial^\mu\partial^\nu K + h.c.
\end{equation}
and the propagator from \cite{Dav}:
\begin{eqnarray}
S_{\mu\nu,\mu^\prime\nu^\prime}&=&\frac{\not q+\sqrt{s}}{10(q^2-M^2+iM\Gamma)}\nonumber\\ &\times&( 5P_{\mu\mu^\prime}P_{\nu\nu^\prime}- 2P_{\mu\nu}P_{\mu^\prime\nu^\prime}+5P_{\mu\nu^\prime}P_{\nu\mu^\prime}\nonumber\\
&+&P_{\mu\rho}\gamma^\rho\gamma^\sigma P_{\sigma\mu^\prime}P_{\nu\nu^\prime}
+P_{\nu\rho}\gamma^\rho\gamma^\sigma P_{\sigma\nu^\prime}P_{\mu\mu^\prime}\nonumber\\
&+&P_{\mu\rho}\gamma^\rho\gamma^\sigma P_{\sigma\nu^\prime}P_{\nu\mu^\prime}
+P_{\nu\rho}\gamma^\rho\gamma^\sigma P_{\sigma\mu^\prime}P_{\mu\nu^\prime}
) ,
\end{eqnarray}
which obeys the criterion formulated above. Here $P_{\mu\nu}=-g_{\mu\nu}+q_\mu q_\nu/s$.

We construct the effective Lagrangian for a 7/2 spin similarly to those for lower spins:
\begin{equation}
L=\frac{f_{BYK}}{m_\pi^3}\overline{B}\left\{1\atop\gamma_5\right\} Y_{\mu\nu\rho}\partial^\mu\partial^\nu\partial^\rho K + h.c.
\end{equation}
An explicit expression for the $7/2$ propagator were proposed recently \cite{Chin}. We obtained and used another expression, which is displayed and commented in Appendix.

Apart from the ingredients discussed above, a form factor suppressing high-momentum contributions is needed, which is especially important for high-spin exchanges. We employ the simple Gaussian form
\begin{equation}
F(q)=\exp(-\vec{q}^2/\Lambda^2)\label{eqff}
\end{equation}
used in \cite{Rij} within another approach. Here $\vec{q}$ is the c.m. 3-momentum of $\overline{K}$ or $K$. This form factor is involved to each vertex in the diagrams of fig. 1. 

Form (\ref{eqff}) differs from monopole, dipole, and other widely applied forms. We examined also other forms and found that form factor (\ref{eqff}) is most successful for the data description. It reduces the cross sections at higher energies stronger than other parametrizations, which is compatible with the data.

Evidently, $F(q)$ is the function of $s$. Strictly speaking, this factor cannot be considered as a vertex form factor for the $u$ channel exchange diagrams. We adopt the same factor for the $u$ channel amplitudes too considering that form factors in such models actually are phenomenological ingredients rather than fundamental quantities. Similar practice has already been used in the literature within other frameworks ({\it e.g.}, \cite{Rij,SLee}).

\section{Input data}
\label{s3}
Data on the integral and differential cross sections and polarizations were published \cite{Bell,Berg,Brie,Brie2,Burg,Carl,Carm,Daub,Gris,Haque,London,Rader,Trip,Tro,Sche} several decades ago. The integral cross sections are compiled in \cite{Flam}. They have been displayed in various ways. Some of them have been tabulated, some of them have been shown on graphs, and others have been presented as parametrizations (Legendre polynomial expansions). We accept only the cross sections presented directly (digitally or graphically) for our fitting procedure. As for Legendre polynomial parametrizations, they determine the cross sections (and polarization) with uncertainties, which magnitude is not always clear. In some cases they give even unphysical (negative) cross sections in limited angular ranges. 
Therefore, we do not deal with the Legendre parametrizations in the fitting procedure. However, we check consistency of our results with the data presented in this form (see the next section).

Most of the data used has been presented with experimental error bars. (Note that, reading the data and, especially, errors from graphs, we possibly introduce small additional uncertainties.) When errors have not been displayed, we simply estimate the statistical errors as $\sqrt{N}$, where $N$ is the number of events.

To fit our model parameters, we use the data as follows:
\begin{itemize}
	\item {integral cross sections of reaction (\ref{e1}), 56 points \cite{Bell,Berg,Brie,Burg,Carm,Daub,Gris,Haque,London,Rader,Trip,Tro,Sche};}
	\item {differential cross section of reaction (\ref{e1}) at 11 values of c.m. energy (1.95, 1.97, 2.07, 2.11, 2.14, 2.24, 2.28, 2.33, 2.42, 2.48, and 2.79 GeV), 234 points \cite{Burg,Daub,Haque,London,Trip,Tro};}
	\item{integral cross sections of reaction (\ref{e2}), 30 points \cite{Bell,Berg,Brie,Carl,Daub,Sche};}
	\item{differential cross section of reaction (\ref{e2}) at 8 values of c.m. energy (1.97, 2.02, 2.07, 2.11, 2.14, 2.15, 2.28, and 2.47 GeV), 76 points \cite{Berg,Burg,Carl,Daub};}
\end{itemize}
totally 396 points.

\begin{table*}
\caption{Results of the fits described in the text: $\chi^2$ values, products of coupling constants (\protect\ref{fff}), and cutoff parameters ($\Lambda_1$ for the low-mass particles and $\Lambda_2$ for high-mass ones).}
\label{tab}       
\begin{tabular}{lccccccccc}
\hline
& $\chi^2$ & $F_\Lambda$ & $F_\Sigma$ & $F_{\Sigma(1385)}$ & $F_{\Lambda(1520)}$ & $\Lambda_1$ (MeV) & $F_{\Sigma(2030)}$ & $F_{\Sigma(2250)}$ & $\Lambda_2$ (MeV)\\
\hline
Without high-mass resonances & 1121 & 0.4502 & 0.1766 & 0.0340 & $-0.6462$ & 773 &&&\\ With high-mass resonances & 985 & 0.3303 & 0.1185 & $-0.0057$ & $-0.4076$ & 839 & 0.0203 & $-0.0838$ & 440\\
\hline
\end{tabular}
\end{table*}

Also a very limited sample of data on reaction (\ref{e3}) is available \cite{Berg,Brie2,Sche}. Since they influence the fitting results rather weakly we do not include them into the fitting procedure. Instead, we use these data for independent checking of our model.

Similarly, we do not include existing data on $\Xi$ polarization \cite{Daub,Trip} to the fit. The rather crude polarization data also have almost no effect on the fitting results. We discuss the polarization data and their description in our model in subsect.\ \ref{s4b}.

\section{Results and discussion}
\label{s4}
\subsection{Fit with low-mass exchange particles}
\label{s4a}
In the first stage, we restrict ourselves to four low-mass exchange hyperons in the $s$ and $u$ channels. In this stage, we have five free parameters: four products of the coupling constants 
\begin{equation}
F_Y\equiv f_{NYK}f_{\Xi YK}\label{fff}
\end{equation}
and one universal cutoff parameter $\Lambda_1$.  All these parameters are determined by the direct fitting to the data.

\begin{figure*}
\includegraphics*{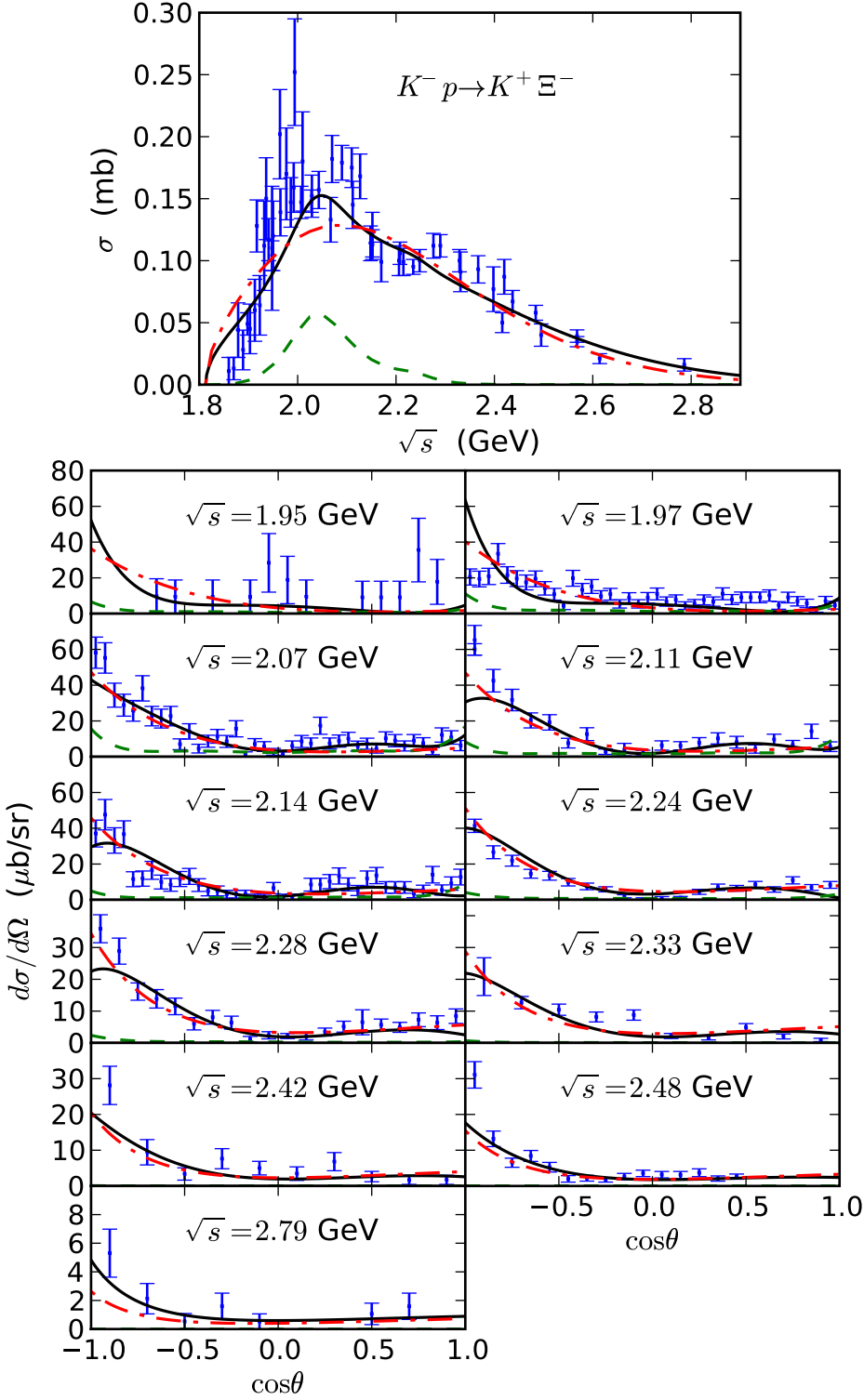}
\caption{Integral (top panel) and differential (other panels) cross sections of the reaction (\protect\ref{e1}). The solid lines are the results from the full version of the model. The cross sections calculated only with the $s$ channel exchanges by $\Sigma(2030)$ and $\Sigma(2250)$ are shown by dashed lines. At $\sqrt{s}>2.4$ GeV, the resonance contribution becomes negligible and dashed lines cease to be visible. The dot-dashed lines are obtained from the fit without higher-mass resonances. The experimental differential cross sections are taken from \protect\cite{Tro} (1.95 GeV), \protect\cite{Burg} (1.97, 2.07, and 2.14 GeV), \protect\cite{Daub} (2.11, 2.28, 2.42, and 2.48 GeV), \protect\cite{Trip} (2.24 GeV), \protect\cite{London} (2.33 GeV), and \protect\cite{Haque} (2.79 GeV).}
\label{f2}
\end{figure*}

\begin{figure*}
\includegraphics*{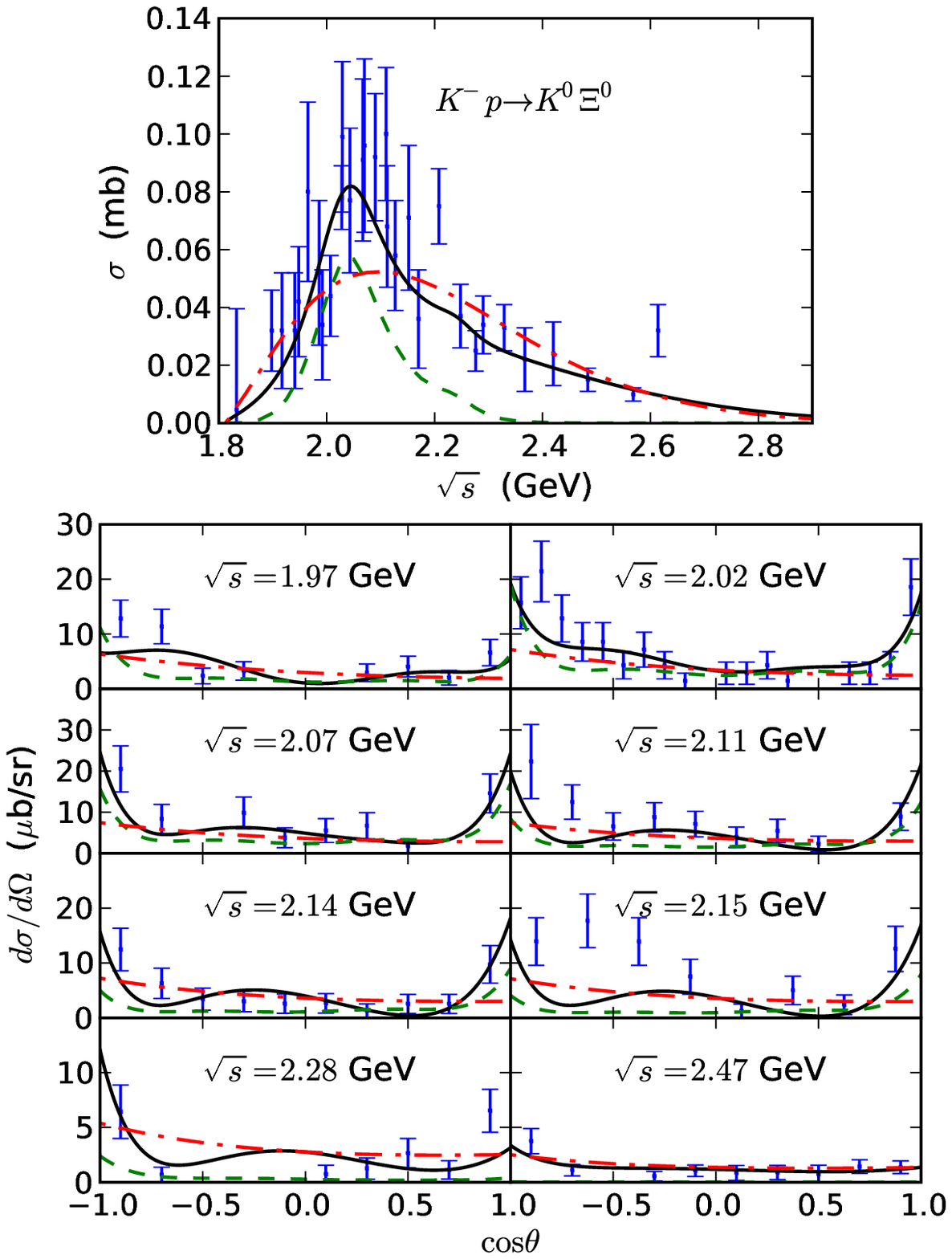}
\caption{The same as in fig. \protect\ref{f2} for reaction (\ref{e2}). The experimental differential cross sections are taken from \protect\cite{Burg} (1.97, 2.07, and 2.14 GeV), \protect\cite{Berg} (2.02 GeV), \protect\cite{Daub} (2.11, 2.28, and 2.47 GeV), and \protect\cite{Carl} (2.15 GeV).}
\label{f3}
\end{figure*}

We performed the fit with four exchange particles: $\Lambda(1/2^+)$, $\Sigma(1/2^+)$, $\Sigma(1385,3/2^+)$, and $\Lambda(1520,3/2^-)$ in the $s$ and $u$ channels. All these particles lie well below the threshold (1812--1819 MeV depending on the charge channel). Comparison of the model and experimental data is presented in figs.\ \ref{f2} and \ref{f3} for reactions (\ref{e1}) and (\ref{e2}), respectively. Results of this fit are shown in the figures by dot-dashed lines. The parameters obtained and the $\chi^2$ value are shown in table \ref{tab}. 

We tried to vary the number of low-lying exchange particles. First, even the simple version with only two exchange hyperons ($\Lambda$ and $\Sigma$) reproduces the data qualitatively. However, the quantitative agreement is poorer ($\chi^2=1683$), particularly, the backward peak in the differential cross sections is strongly and systematically underestimated.

On the other hand, when we add the fifth exchange particle, $\Lambda(1405)$, to the four ones, we do not get a meaningful improvement. Hence, we conclude that the four-particle scheme is practically optimal as the starting point of our model. Of course, it should be noted that the model is constructed on purely phenomenological grounds. We do not think that exchanges by other particles are truly absent. Instead, the four exchanges reflect effectively a more complex real physical picture.

It is seen from figs.\ \ref{f2} and \ref{f3} that the fit reproduces well the main features of the cross sections. The energy dependence of the integral cross sections is described well within the whole energy range with a physically reasonable value of  cutoff parameter $\Lambda_1$. The backward peak in the differential cross sections is fairly reproduced though some underestimation at certain energies should be also noticed. Shape evolution of the differential cross sections with energy is also reproduced.

We checked a possible contribution of the $t$ channel exchange mechanism adding the diagram with one-boson exchange by a double-strangeness meson. This meson, of course, need not be assigned to any non-$\overline{q}q$ exotics, but rather may be treated as a simulation of two-kaon exchanges. Varying reasonably properties of this effective meson, we did not obtain essential improvement. So we suppose that the $t$ channel exchange plays a minor role.

Obtained value $\chi^2=1121$ at 391 degrees of freedom is not low. In our opinion, an important contribution to this value originates from experimental pitfalls. Compare, for instance, the experimental cross sections at energies very close to each other, $\sqrt{s}=2.14$ and 2.15 GeV, from different experiments \cite{Burg} and \cite{Carl}, respectively, in fig.\ \ref{f3}. It is seen that they differ from each other strongly. Physically, such behavior can be explained only if a very narrow resonance exists in this region, which is rather unlikely. Probably, no realistic theory can reproduce these cross sections simultaneously. Thus, one can suggest that the data are uncertain. Some other similar though less prominent self-contradictions of the data can be indicated. So we think that the $\chi^2$ value can not be reduced radically. Of course, this restricts possibilities to deduce a fully reliable theoretical picture.

On the other hand, some drawbacks of our description appearing systematically should be noted. The integral cross sections in the bump region (approximately 1.9--2.1 GeV) are underestimated. We can not describe the forward peak, which is seen in reaction (\ref{e2}) [but not in reaction (\ref{e1})]. Unfortunately, this peak is concentrated in the single bin at each energy, so its structure remains unclear. Possibly, our model in the four-particle version can not describe also the data on $\Xi$ polarization (see the next subsection). We tried various modifications of the model restricted to sub-threshold exchange particles. Particularly, we examined the form factor depending on $u$ in the $u$ channel diagrams, which can affect the angular distribution. We employed the form from ref. \cite{Nak}. However, in this and some other ways we could not achieve essential improvement. So, we suggest that feasibility of the model with sub-threshold exchange particles is limited. An extension of the model is presented in the next subsection.

\subsection{Incorporation of above-threshold resonances}
\label{s4b}
As mentioned above, eight well established $Y$ resonances in the relevant energy range are known (and also more than ten doubtful ones). Of course, it is impracticable to incorporate all of them in the calculation since the available data do not give a possibility to determine eight additional parameters (products of coupling constants). Moreover, masses and, especially, widths of the resonances are also uncertain and, strictly speaking, may be free parameters too.

We examined various combinations of several resonances from the eight known ones adding to the four-particle scheme as above and obtained in several cases substantial improvement of the data description. Ultimately, we choose $\Sigma(2030)$ and $\Sigma(2250)$ stressing that this choice is not unique. Spin and parity ($J^\pi =7/2^+$) of $\Sigma(2030)$ are well established. For $\Sigma(2250)$, most probable assignments are $5/2^-$ and $9/2^-$ \cite{PDG}. We choose $J^\pi =5/2^-$ for $\Sigma(2250)$ checking that other choices do not change the results drastically. We fix the ``nominal'' masses (2030 and 2250 MeV) and widths 175 and 105 MeV, respectively. The latter quantities are the median values of the ranges presented for the widths in the PDG compilation \cite{PDG}.

The corresponding form factor is chosen in the same form (\ref{eqff}). However, we need to employ another cutoff parameter $\Lambda_2$ for the above-threshold resonances. The reason is as follows. The fitting requires for the low-mass particles the values of cutoff parameter $\Lambda_1$ about 1 GeV. If we calculate an isolated high-mass and high-spin resonance with such values of $\Lambda_1$, we obtain a peak shifted to the right by several hundreds MeV from the nominal mass. We consider such behavior as inconsistent with empirical data because the resonances are revealed as peaks at near-nominal masses. More hard cutoffs ($\Lambda_2$ about 0.5 GeV) put peaks to their proper places.

\begin{figure*}
\includegraphics*{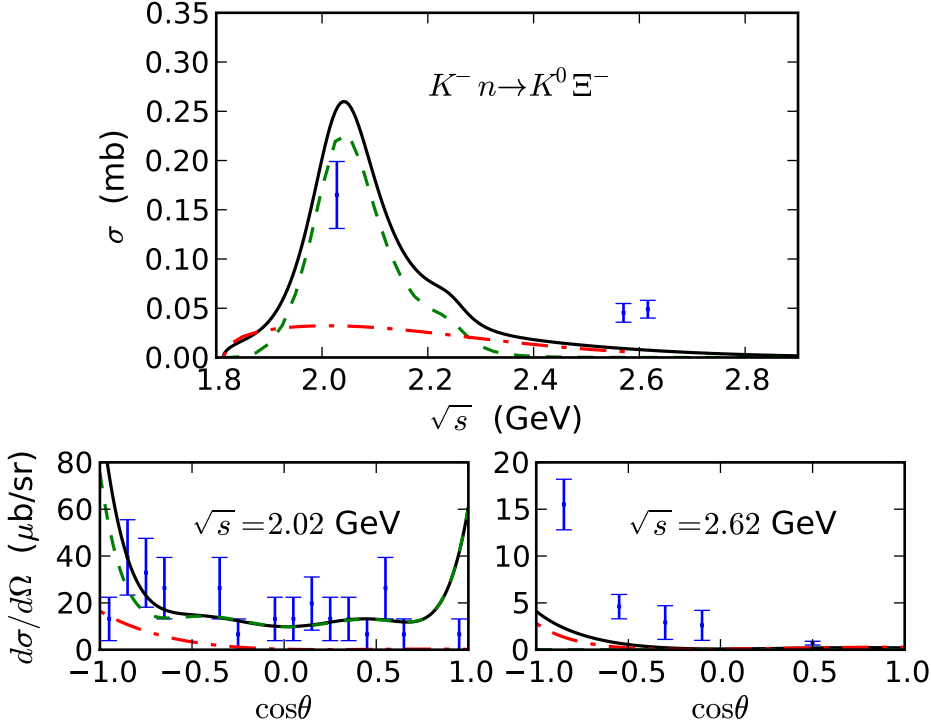}
\caption{The same as in fig. \protect\ref{f2} for reaction (\ref{e3}). The experimental differential cross sections are taken from \protect\cite{Berg} (2.02 GeV) and \protect\cite{Sche} (2.62 GeV).}
\label{f4}
\end{figure*}

Results of our final fit are shown in figs.\ \ref{f2} and \ref{f3} by solid lines. The best fit parameters are presented in table \ref{tab}. It is seen that the description of the data is improved substantially though $\chi^2$ decrease is not so large. Particularly, the drawbacks mentioned in the previous subsection are remedied. First, a strong forward peak appears for reaction (2) in agreement with the experiments. The bump in the integral cross sections is described adequately now.

The data manifest prominent forward peaks at $\sqrt{s}=2.0-2.3$ GeV in the differential cross sections of reaction (2), but not (1). Our analysis gives a simple explanation of this feature. It appears that reaction (2) in the bump region is almost purely resonant (see dashed lines in fig. \ref{f3}). The higher-mass resonances give nearly equal contribution to the cross sections of both reactions. However, reaction (1) is also substantially contributed by $\Lambda$ and $\Lambda(1520)$ $u$ channel exchanges, which are impossible for reaction (2). That is why the differential cross section of reaction (2) is more symmetric than that of reaction (1) and the ``bump'' integral cross section of reaction (1) is about twice as large as that of reaction (2).

The integral and differential cross sections of reactions (1) and (2) considered as most reliable are used for the fitting. Besides, other data mentioned in sect. \ref{s3} exist and they are not included to the fitting. It should be noted that the total statistical weight of those data is relatively small so they anyway cannot affect the fitting strongly, considerably increasing amount of the calculations in some cases. We prefer to use these data for independent checking of our model.

In fig.\ \ref{f4}, the integral and differential cross sections of reaction (3) are displayed. It is seen that the addition of the higher-mass resonances improves the description considerably though accurate quantitative agreement is not reached. It is notable that just $\Sigma$ resonances are needed for this improvement while $\Lambda$ resonances in the $s$ channel do not contribute to reaction (\ref{e3}).

\begin{figure*}
\includegraphics*{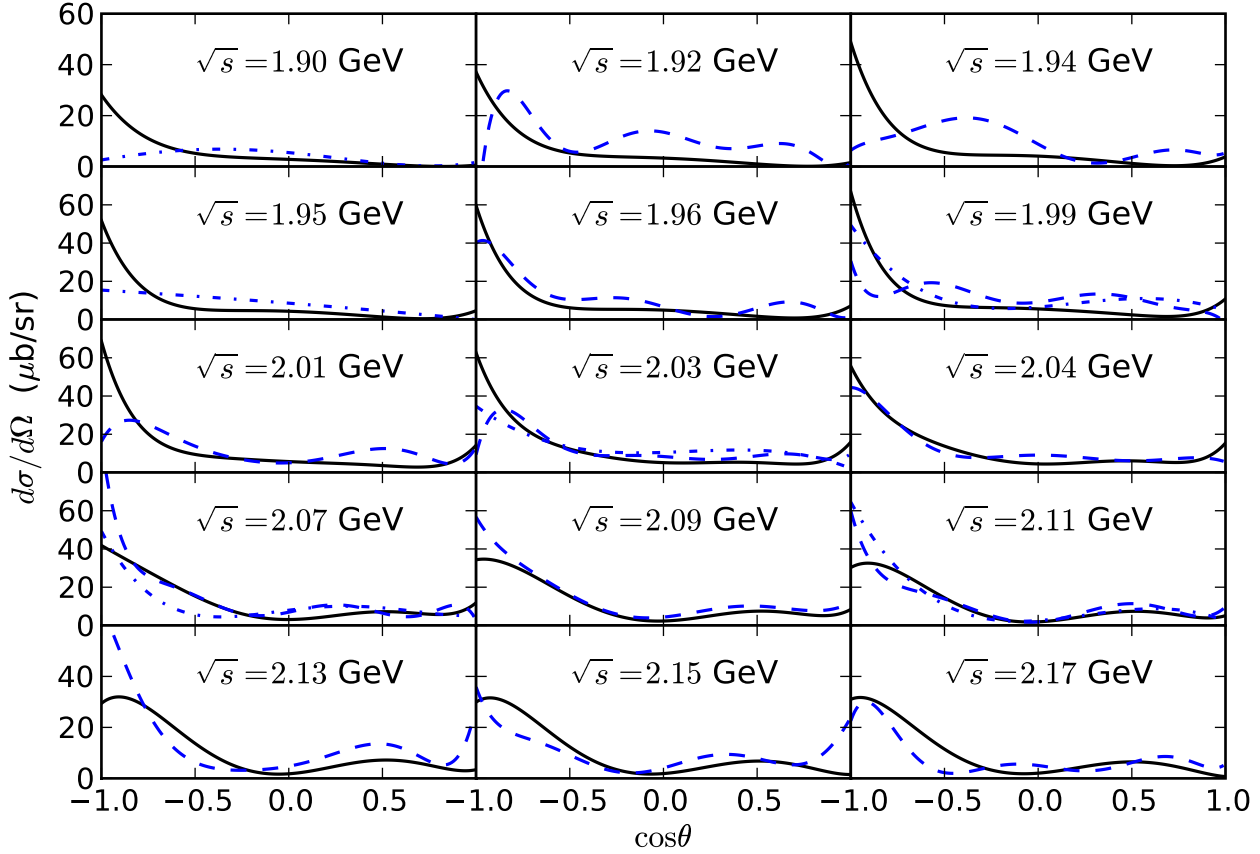}
\caption{Comparison of our calculations (solid lines) with the experimental differential cross sections of the reaction (\protect\ref{e1}) restored from the Legendre expansions \protect\cite{Burg} (dashed) and \protect\cite{Berg} (dot-dashed).}
\label{f5}
\end{figure*}

Then, we show in fig.\ \ref{f5} the differential cross sections restored from the Legendre polynomial expansions compared to our calculations. Since the uncertainties of these data are not well defined, the agreement can be estimated only visually. It is seen that the theory is reasonably consistent with the experiment. Considerable discrepancies at the near-threshold energies are possibly associated with final-state interactions neglected in our approach.

\begin{figure}
\includegraphics*{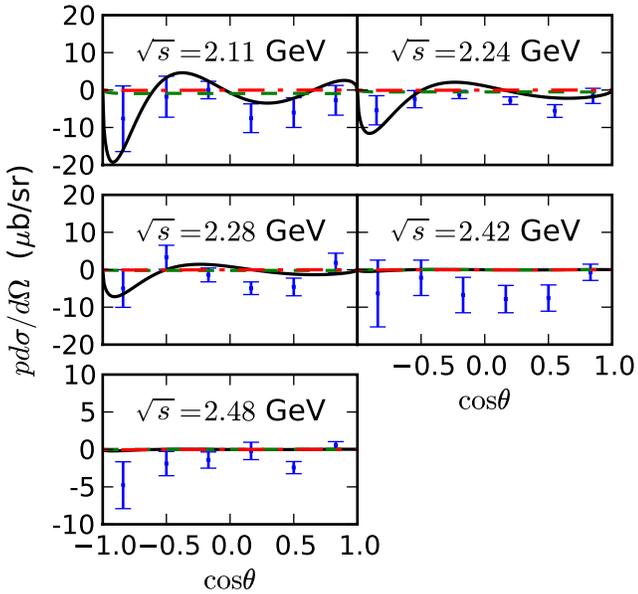}
\caption{Polarization of $\Xi$ hyperons times the differential cross section from reaction (\protect\ref{e1}). Convention for curves is the same as in fig. \ref{f2}. The experimental data are taken from \protect\cite{Trip} (2.24 GeV) and \protect\cite{Daub} (the other energies).}
\label{f6}
\end{figure}

\begin{figure}
\includegraphics*{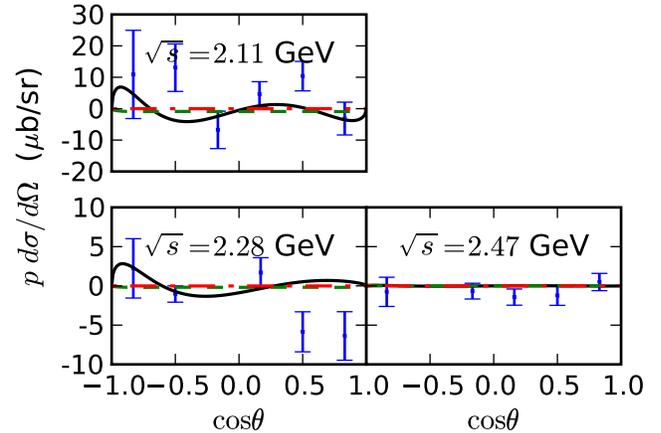}
\caption{The same as in fig. \protect\ref{f6} for reaction (\protect\ref{e2}). The experimental data are taken from \protect\cite{Daub}.}
\label{f7}
\end{figure}

Another sample of data is associated with $\Xi$ hyperon polarization. All the data that can be directly taken from papers \cite{Daub,Trip} are presented in figs.\ \ref{f6} and \ref{f7}.

Without the above-threshold resonances, the polarization in our model is practically zero. Since we work within the Born approximation, nontrivial imaginarities in the amplitudes leading to nonzero polarization appear only from the widths in the propagator denominators. The subthreshold resonances have relatively small widths and can not give a significant polarization themselves. A sizable polarization appears from the interference between the higher-mass and lower-mass exchanges. Note that at $\sqrt{s}>2.4$ GeV (beyond the resonance region) the polarization becomes almost zero again.

The data are crude and even the hypothesis of exact zero polarization cannot be rejected. However, it seems that the final version of the model gives a better correspondence with the data. We consider this as one more argument for the important role of the higher-mass resonances though it is possible that effects neglected in our model (like initial and final state interactions) can change the picture substantially.

\begin{figure*}
\includegraphics*{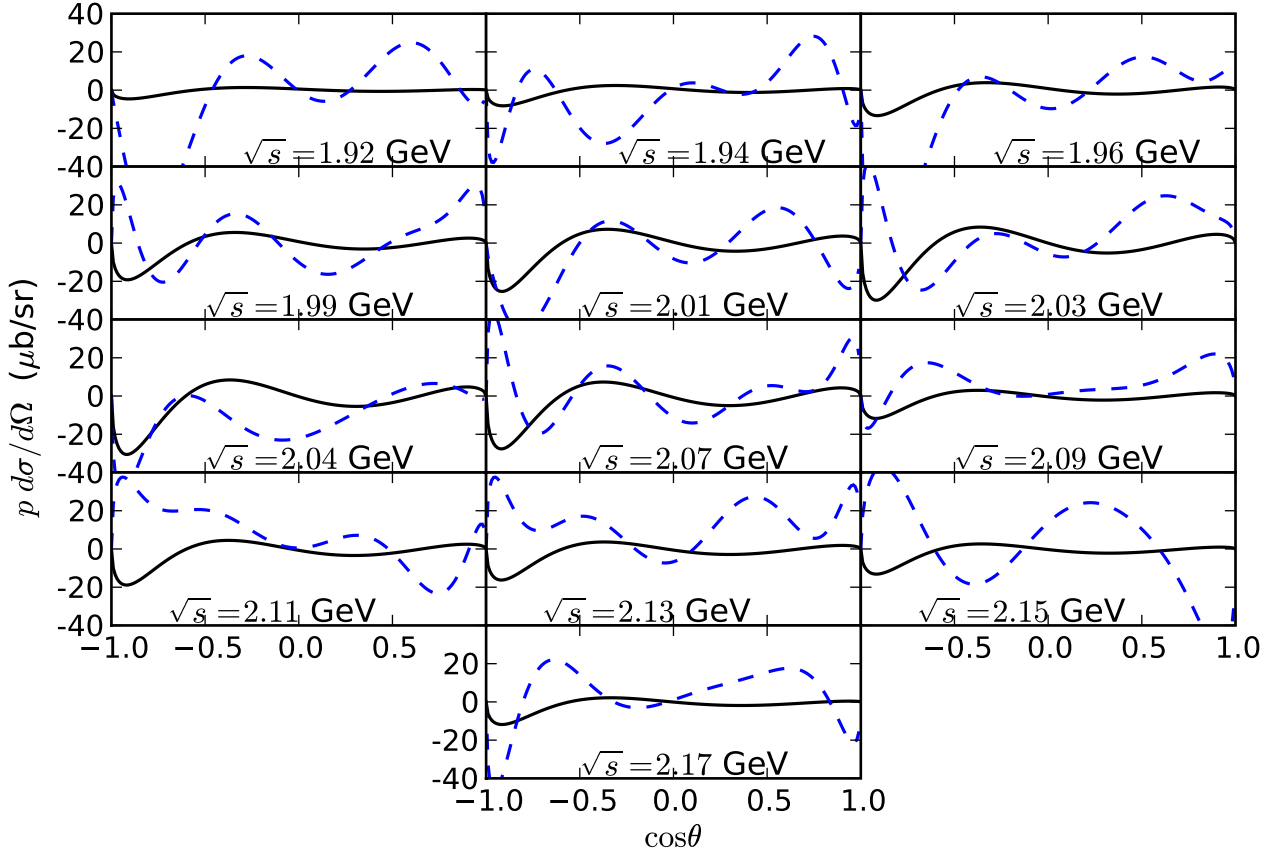}
\caption{Polarization of $\Xi$ hyperons times the differential cross section from reaction (\protect\ref{e1}) restored from the Legendre expansions \protect\cite{Burg} (dashed) compared with our calculations (solid).}
\label{f8}
\end{figure*}

In fig.\ \ref{f8}, the polarization data restored from the polynomial expansions \cite{Burg} are presented together with our calculations. Uncertainties of the data are not well defined again, so only visual comparison is possible. It is difficult to deduce robust conclusions from this comparison. Note that the data are often in contradiction not only with the calculations, but also with the data at a very close energy (compare, for instance, the data at $\sqrt{s}=1.99$ and 2.01, or 2.15 and 2.17 GeV). The largest discrepancies are seen in the region of backward angles where the polarization varies very sharply. (Note that the polarization at $\theta=0^0$ and $180^0$ is, of course, exact zero, while the curves are sometimes almost vertical in these regions so it may seem visually that the polarization has some nonzero value.)

As mentioned above, we tested also combinations of the higher-mass resonances other than $\Sigma(2030)$ and $\Sigma(2250)$. Here we explain our final choice.

Underestimation of the integral cross sections in the bump region pointed out in subsect. 4.1 requires a resonance with the mass about 2.0--2.1 GeV. We observed that $\Lambda(2100)$, $\Lambda(2110)$, and $\Sigma(2030)$ give similar quality of description of the integral cross sections of the reactions (\ref{e1}) and (\ref{e2}). Our choice of $\Sigma(2030)$ is inspired by the reaction (\ref{e3}) (see fig. 4), where only $\Sigma$ resonances can contribute in the $s$ channel. It should be noted, however, that there is only one experimental point in the bump region for the integral cross section of this reaction.

The significant role of the $\Sigma(2030)$ resonance in the $\overline{K}N\to K\Xi$ reaction was suggested by us in \cite{KLSh}. It is remarkable that very recently the similar conclusion was drawn \cite{Man} for $\Xi$ photoproduction $\gamma p\to K^+K^+\Xi^-$.

The second resonance, $\Sigma(2250)$, provides a somewhat better description of the cross sections at the right slope of the bump. Also substantially nonzero polarization appears in a wider energy range. Since the spin of $\Sigma(2250)$ is not yet established, we checked the sensitivity of the results to the spin value. Changing the spin to 7/2 and 9/2, we found only small modifications of the results, though the fit became a bit worse.

We tried also to incorporate larger number of the higher-mass resonances (up to seven). Of course, the fit becomes more ambiguous and sometimes multiple minima of $\chi^2$ appear. More important, we did not find great benefits in the data description. So we conclude that the suggested model is adequate to the existing amount of the data.

We calculated the partial widths of $\Sigma(2030)$ and $\Sigma(2250)$ decays into $K\Xi$ using our products of the vertex constants and known branching ratios for the $\overline{K}N$ channel from \cite{PDG}. We obtained 1.2 and 0.3 MeV, respectively. Though the total widths are known with large ambiguities, we can roughly estimate the corresponding branching ratios as 0.7\% and 0.3\%, respectively, which is not in contradiction with small upper limits deduced in \cite{PDG}. It is remarkable that so small branching ratios imply in the cross sections significantly.

Many authors tried to extract vertex constants for couplings $NK\Lambda$ and $NK\Sigma$ (for a recent compilation, see \cite{GC}). Some predictions for the $\Xi K\Lambda$ and $\Xi K\Sigma$ couplings are made within general $SU(3)$ schemes ({\it e.g.}, \cite{OBE1}) or in more specialized approaches \cite{Choe,Zamir}. Some fragmentary attempts to derive the constants for resonances are also known ({\it e.g.}, \cite{Rij,Nak}).

Even for the $NK\Lambda$ and $NK\Sigma$ couplings, the values deduced by various ways are several times different \cite{GC}. Often only the coupling constants are presented rather than the full vertex functions while the coupling constants practically have a physical meaning only together with the relative form factor. For this reason, the comparison of our constants with highly ambiguous and widely scattered corresponding quantities from the literature is not instructive. Note that we determine only products of the coupling constants, and our form factors are not normalized to unity on the mass shell as seen from (\ref{eqff}).

We examined the set of the coupling constants from \cite{Nak}. Incorporating also form factors from \cite{Nak} and treating the cutoff parameter as a free one, we can not achieve quantitative agreement with the data. The salient difference between the two models is considerably stronger couplings $F_{\Sigma(1385)}$ and $F_{\Lambda(1520)}$ in \cite{Nak} with respect to our ones. To compare properly the couplings, we fixed the constants from \cite{Nak} and fit parameters $\Lambda$ of our form factor (\ref{eqff}) allowing the cutoff parameters to be different for different hyperons. We obtain rather hard cutoffs ($\Lambda=400-500$ MeV) for $\Sigma(1385)$ and $\Lambda(1520)$ contrary to the ground state $\Lambda$ and $\Sigma$. This means that the data on the reaction 
$\overline{K}N\to K\Xi$ require to reduce these vertex functions. Note that our constants are purely phenomenological while the constants in \cite{Nak} are determined from $SU(3)$ symmetry relations.

Reaction $(K^-,K^+)$ on nuclei becomes a hot topic now owing to the study of $\Xi$ hypernuclei starting at J-PARC \cite{Nagae,Nagae2}. Predictions of the cross sections of the $^A$Z$(K^-,K^+){}^A_\Xi($Z$-2)$ reaction ({\it e.g.}, \cite{Mot}) incorporate proton effective numbers, which are determined by the nuclear and hypernuclear structure, and properties of elementary process (\ref{e1}).

\begin{figure}
\includegraphics*{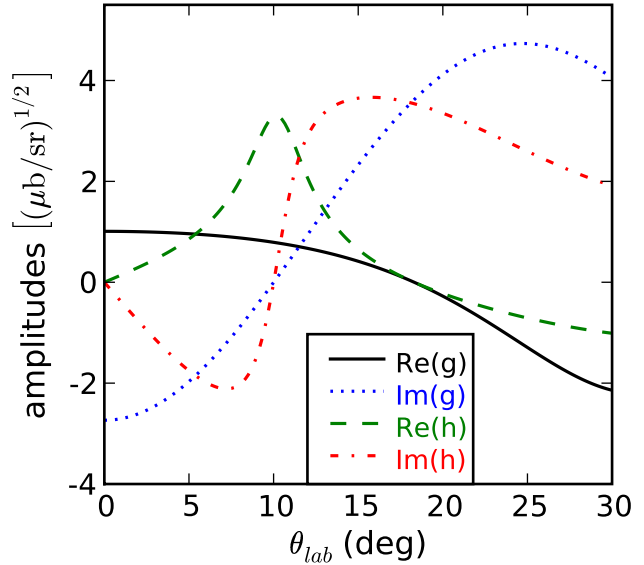}
\caption{Spin-independent and spin-flip amplitudes at $k_{lab}=1.8$ GeV/c ($\sqrt{s}=2.14$ GeV) in the laboratory system at small angles.}
\label{f10}
\end{figure}

Generally, the amplitude of the elementary process has the form
\begin{equation}
f(\theta)=g(\theta)+h(\theta)\vec{\sigma}\cdot \vec{n}.\label{fgh}
\end{equation}
To our knowledge, all calculations for $\Xi$ hypernuclear production so far neglected spin-flip component $h(\theta)$ (with the single exception of ref. \cite{MT} where the $h$ amplitude is chosen purely arbitrarily). In this case, the single input related to the elementary process is the differential cross section, which can be taken \cite{Mot} directly from the data.

The spin-flip component can lead to some important effects in hypernuclear production, for instance, to abnormal parity transitions as was considered earlier \cite{Mot2} for $\Lambda$ hypernuclei. In view of hypernuclear applications, we present in fig.\ \ref{f10} amplitudes $g$ and $h$ in the laboratory system at incident momentum 1.8 GeV/c and forward angles. It is seen that spin-flip amplitude $h$ grows quickly with the angle and becomes comparable with $g$ even at $\theta_{lab}\sim 5^0$. Therefore, the spin-flip amplitude is generally non-negligible. Probably, $g$ plays nevertheless a leading role, but a study of production of $\Xi$ hypernuclei with taking into account spin-flip effects is worthwhile.

\section{\label{s5}Conclusion and outlook}

We have formulated the phenomenological model for the $\overline{K}N\to K\Xi$ process. While binary reactions for production of $\Lambda$ and $\Sigma$ hyperons are studied thoroughly, this process was completely out of theoretical scope for the last decades. Though the quality of the data is far from adequate, dealing with the whole set provides a possibility of the quantitative analysis of the reaction.

Our goal was to elaborate a working tool for treating the reaction of $\Xi$ production rather than to consider it from the fundamental point of view. Therefore, the model should be regarded as effective one. We tried to minimize the number of exchange particles so the coupling constants possibly include effectively contributions of other hyperons left aside the model. Nevertheless, we hope that the model can be a sizable step for developing the theory of more complicated reactions of $\Xi$ production, particularly, photoproduction $\gamma p\to K^+K^+\Xi^-$. The model suggested in \cite{Nak} for the photoproduction fails to describe the data on the $\overline{K}N\to K\Xi$ reaction. This is not so surprizing because the parameters of the model of ref. \cite{Nak} as well as the parameters of our model are effective and probably incorporate implicitly different neglected effects. Consistent simultaneous treatment of various reactions of $\Xi$ production is an ambitious future task.

Earlier, $u$ channel exchange was considered as the dominant mechanism of the $\overline{K}N\to K\Xi$ reaction. Confirming an important role of the $u$ channel exchange, we show that the contribution of the on-shell resonances is also significant. We cannot claim that just $\Sigma(2030)$ and $\Sigma(2250)$ work (other choices are also possible), but establish that taking into account of the resonance mechanism is necessary. It is notable that relatively small (not higher than 1\%) branching ratios of $Y\to K\Xi$ decays are manifested in the cross sections (which are not large themselves) clearly. We show that charge channel $K^-p\to K^0\Xi^0$ is ``more resonant'' than $K^-p\to K^+\Xi^-$ due to lack of $u$ channel $\Lambda$ exchange in the former case. Maybe, in future this feature can be used for delicate studies of high-lying $Y$ resonances in view of, for instance, J-PARC capabilities.

Since the threshold of the reaction is relatively high, the majority of relevant resonances has large spins. Hence, we presented here technicalities for spin 7/2 exchanges, which can be helpful also for other reactions.

As usual, polarization is the quantity rather sensitive to reaction mechanisms. Seemingly, our model catches the main trends of the polarization though more accurate data on polarization (as well as the cross sections) are needed for robust conclusions.

The model is directly applicable to the theory of production of $\Xi$ hypernuclei. While empirical data on the elementary process do not enable one to extract certainly the spin-flip amplitude, the model do. Predictions of hypernuclear spectra measurable in the $(K^-,K^+)$ reaction with account of the spin structure of the elementary amplitude and analysis of coming data from J-PARC is another task for the future.

\begin{acknowledgement}
This work was supported in part by Russian Foundation for Basic Research, grant No.\ 08-02-00510.
\end{acknowledgement}

\section*{Appendix}
\renewcommand{\theequation}{A.\arabic{equation}}
\setcounter{equation}{0}
General recipes for propagators of particles with arbitrary spins were elaborated long ago \cite{BF,Fron,Wein,Scad}. While the propagator for spin 5/2 also has long been known \cite{RR}, the 7/2 case was treated explicitly \cite{Chin} much later. Huang {\it et al.} \cite{Chin} using the prescription of Behrends and Fronsdal \cite{BF,Fron} first derived the expression for the spin 4 projection operator $P_{\mu_1\mu_2\mu_3\mu_4,\nu_1\nu_2\nu_3\nu_4}(4)$ and then got the 7/2 propagator as
\begin{equation}
S_{\mu_1\mu_2\mu_3,\nu_1\nu_2\nu_3}=\frac{\not q+\sqrt{s}}{q^2-M^2+iM\Gamma}\frac 49 \gamma^\alpha\gamma^\beta P_{\alpha\mu_1\mu_2\mu_3,\beta\nu_1\nu_2\nu_3}(4).\label{bf}
\end{equation}

We repeated their derivation and obtained the following expression:
\begin{eqnarray}
S_{\mu_1\mu_2\mu_3,\nu_1\nu_2\nu_3}&=&\frac{\not q+\sqrt{s}}{q^2-M^2+iM\Gamma} \nonumber\\ &\times&\left( \frac 16 \sum_{\{\nu\}} P_{\mu_1\nu_1}P_{\mu_2\nu_2}P_{\mu_3\nu_3}\right. \nonumber\\ &&-\frac 1{84}\sum_{\{\mu\nu\}} P_{\mu_1\mu_2}P_{\nu_1\nu_2}P_{\mu_3\nu_3}
\nonumber\\ &&-\frac 1{84}\sum_{\{\mu\nu\}} P_{\mu_1\nu_1}P_{\mu_2\nu_2}R_{\mu_3\nu_3}
\nonumber\\ &&\left. +\frac 1{420}\sum_{\{\mu\nu\}} P_{\mu_1\mu_2}P_{\nu_1\nu_2}R_{\mu_3\nu_3}\right).
\label{72}
\end{eqnarray}
Here
\begin{equation}
P_{\mu\nu}=-g_{\mu\nu}+\frac{q_\mu q_\nu}{s}\label{AP}
\end{equation}
and
\begin{equation}
R_{\mu\nu}=-\gamma_\mu\gamma_\nu-\frac{\gamma_\mu q_\nu-\gamma_\nu q_\mu}{\sqrt{s}}+\frac{q_\mu q_\nu}{s}.\label{AR}
\end{equation}

The sums in (\ref{72}) are taken over all permutations of $\nu$ indices in the first sum (6 terms) and over all permutations of $\mu$ as well as $\nu$ indices in the other sums (36 terms each). Due to simple identities $P_{\alpha\beta}=P_{\beta\alpha}$ and $P_{\alpha\beta}P_{\gamma\delta}=P_{\gamma\delta}P_{\alpha\beta}$, the second, third, and fourth sum contains 9, 18, and 9 nontrivial terms, respectively.

The final expression from \cite{Chin} can be simplified and reduced to the same form as (\ref{72}). However, factor $-103/7560$ appears in front of the second sum instead of our $-1/84$. We checked numerically that (\ref{bf}) and (\ref{72}) (but not the formula from \cite{Chin}) are equivalent. We also checked that propagator (\ref{72}) satisfies the criterion from sect. \ref{s2} (gives correct angular distribution) while the propagator from \cite{Chin} does not. So we believe that expression (\ref{72}) is correct.

Of course, one may use directly eq. (\ref{bf}). But in cumbersome calculations, the explicit formula (\ref{72}) is more convenient and time-saving.

For the partial width of the channel $Y(7/2^\pm)\to K(0^-)B(1/2^+)$ we obtain:
\begin{equation}
\Gamma= \frac{f^2_{YKB}}{70\pi m_\pi^6}\frac{E_B\pm M_B}{M_Y}q^7,\label{width}
\end{equation}
where the sign in the numerator coincides with the parity of the resonance, $q$ is the c.m. 3-momentum of decay particles. Estimating the partial width in sect. \ref{s4}, we multiplied (\ref{width}) by the relevant form factor and calculate $q$ at the nominal mass of the resonance.

\end{document}